\documentclass[epj]{svjour}
\usepackage{graphicx}
\usepackage{amssymb}
\usepackage{dcolumn}
\usepackage{amsmath}
\usepackage{bm}
\usepackage{textcomp}
\usepackage{epstopdf}
\usepackage{color,url}
\usepackage{ulem}
\usepackage[toc,page,title,titletoc,header]{appendix}
\usepackage[colorlinks,
            linkcolor=blue,
            anchorcolor=blue,
            citecolor=blue,
            urlcolor=blue
            ]{hyperref}
\usepackage{txfonts}
\usepackage{caption}
\usepackage{lineno}
\usepackage{multirow}
\usepackage{ulem}
\begin{document}

\title{Dynamical and finite-size effects on the criterion of first-order  phase transition}
\author{Lijia Jiang\inst{1,2,3} \thanks{{e-mail:} \href{mailto:lijiajiang@nwu.edu.cn}{lijiajiang@nwu.edu.cn} }
\and Fei Gao\inst{4} \thanks{{e-mail:} \href{mailto:fei.gao@bit.edu.cn}{fei.gao@bit.edu.cn} }
\and Yu-xin Liu\inst{5,6,7} \thanks{{e-mail:} \href{mailto:yxliu@pku.edu.cn}{yxliu@pku.edu.cn} }%
 }

\institute{ School of Physics, Northwest University, Xi'an, 710127, China
\and Shaanxi Key Laboratory for Theoretical Physics Frontiers, Xi'an, 710127, China
\and Peng Huanwu Center for Fundamental Theory, Xi'an 710127, China
\and School of Physics, Beijing Institute of Technology,  100081 Beijing, China
\and Department of Physics and State Key Laboratory of Nuclear Physics and Technology, Peking University, Beijing 100871, China
\and Center for High Energy Physics, Peking University, 100871 Beijing, China
\and Collaborative Innovation Center of Quantum Matter, Beijing 100871, China}

\date{Received: date / Revised version: date}

\abstract{To identify first-order phase transitions in the dynamical process similar to the relativistic heavy-ion collisions, we investigate the dynamical behaviors of the first-order phase transition criterion in the Fokker-Planck framework. In the thermodynamic limit, the criterion can be expressed as combinations of cumulants or coefficients of an Ising-like effective potential. Our study reveals that factors such as phase transition scenarios, initial temperature, system volume, relaxation rate, and evolution trajectory have great impacts on the criterion, a larger initial temperature, a smaller volume, a larger relaxation rate, or bending of the trajectory will all lead to a reduction of the first-order phase transition signal, while volume expansion over time preserves signal integrity. Analysis along a hypothetical freezeout line shows that the signal is possibly preserved at relatively large chemical potentials.}

\maketitle

\section{Introduction}

The properties of quantum chromodynamic (QCD) matter are being explored in the relativistic heavy-ion collisions at the BNL Relativistic Heavy Ion Collider (RHIC). By tuning the collision energy, the QCD matter in the quark-gluon plasma (QGP) state evolves along different isentropic trajectories in the baryon chemical potential ($\mu$) and temperature ($T$) plane, and in this way, the structure of the QCD phase diagram is supposed to be scanned~\cite{Nuclear:2008,STAR:2010vob,Baym:2002,Fukushima:2011,Bzdak:2019pkr,Luo:2020pef,Lovato:2022vgq}.
From the theoretical side, the QCD phase transition at small chemical potentials is confirmed to be crossover by lattice QCD simulations~\cite{Brown:1990,Aoki:2006we,Aoki:2009sc,Karsch:2003jg};
and in the large chemical potential region, it is predicted to be first-order phase transition by nonperturbative QCD approaches like Dyson-Schwinger equations~\cite{Roberts:2000aa,Fischer:2018sdj,Qin:2011,Fischer:2014ata,Gunkel:2021oya,Gao:2020qsj,Gao:2020fbl}, functional renormalization group methods~\cite{Berges:2000ew,Dupuis:2020fhh,Fu:2022gou,Fu:2019hdw,Fu:2021oaw,Fu:2023lcm}, and chiral effective theories such as the holograpic QCD model,  (P)NJL model and the linear sigma model
~\cite{Klevansky:1992qe,Fukushima:2004qe,Fu:2007xc,Herbst:2011,Jiang:2013,Schaefer:2007,Schaefer:20072,Shao:2011fk,He:2013qq,Chelabi:2015gpc,Kojo:2020ztt,Chen:2020ath}.
However, the existence of the first-order phase transition as well as the location of the critical endpoint (CEP) remain to be verified by experiments, and the detection of the QCD phase structure is one of the main goals of the Beam Energy Scan program at RHIC.

The high-order fluctuations of conserved charges have been taken as one of the main experimental observables for the search of the CEP,
since they are related to the increase of correlation length in the critical region~\cite{Stephanov:2008qz,Athanasiou:2010kw,Stephanov:2011pb}.
Experimentally, the collision-energy-dependent cumulant ratio $C_{4}/C_{2}$ ($\kappa\sigma^{2}$) of net-proton production in the Au-Au central collisions presents non-monotonic behavior at medium collision energies, with deviations from both the statistical baselines and the hadron resonance gas model at low collision energies~\cite{STAR:2020tga}.
This feature strongly hints that the phase transition at low collision energies differs from the crossover, still it can not be determined whether it is originated from CEP or first-order phase transition. In addition to the cumulant ratios, the other CEP related experimental observables, such as the yield ratio of light nuclei~\cite{Sun:2018jhg,STAR:2022hbp}, the baryon-strangeness correlation~\cite{Koch:2005vg}, and the  intermittency~\cite{STAR:2023jpm} are also measured at RHIC, all of which exhibit nonmonotonic behaviors and deviations from the statistical or non-critical model simulations around $20$ GeV. To quantitatively explain these experimental data, reliable dynamical modeling and theoretical analysis on the phase transition signals are necessary.
%Further, by fitting its parameters from the experimental data of the first four cumulants, the calculations of the two-component model ~\cite{Bzdak:2018uhv,Bzdak:2018axe} can describe the general trend of the recent experimental data of fifth- and sixth-order fluctuations for net-proton, indicating the possibility of first-order phase transition at low collision energies~\cite{STAR:2022vlo}.

To analyze the occurrence of phase transition in a dynamical system like the fireball expansion at RHIC, kinds of factors, such as the dynamical effects~\cite{Berdnikov:1999ph,Stephanov:2009ra,Mukherjee:2015swa,Jiang:2017mji,Jiang:2017fas,Jiang:2021zla,Sakaida:2017rtj,Nahrgang:2018afz,Jiang:2023nmd,Pihan:2022xcl}, the finite-size effects~\cite{brezin1985npb,Imry1980prb,Spieles:1997ab,Zabrodin:1998vt}, the nonuniform-temperature/baryon chemical potential effects~\cite{Zheng:2021pia},
the late-stage hadronic interactions~\cite{Hammelmann:2023aza}, and the noncritical effects~\cite{Bluhm:2016byc}, etc, should be involved. The dynamical depiction of the critical fluctuations has been gradually developed in recent years.
The dynamical high-order cumulants were first studied in \cite{Mukherjee:2015swa}. Within the Fokker-Planck framework, they presented the importance of dynamical effects on the final-state observables. However, the dynamical cumulants in the first-order phase transition region were not discussed since the discontinuities of the equation of state are not compatible with the differential equations for dynamical evolution. On the other hand, the experimentally generated fireball is a finite system and its typical size is approximately $10^3$ fm$^3$. The discontinuities in the first-order phase transition are rounded due to the finite-size effects, thus the studies of the dynamical high-order cumulants were extended to the first-order phase transition region in \cite{Jiang:2023nmd}. Both the equilibrium and non-equilibrium analysis predict the nonmonotonic behavior of kurtosis approaching the CEP. But how cumulants can be associated with a first-order phase transition signal is rarely discussed.
The measurements on the first-order phase transition signals at RHIC remain challenging~\cite{Nahrgang:2011vn,STAR:2017okv},
and it is quite necessary to theoretical analyze the signals of  first-order phase transition in a dynamic system.

Based on the Landau phase transition theory, the effective potential (the free energy) of the QCD system can be written as a polynomial form of the order parameter. The number of phases is  determined by the coefficients at different physical conditions. In \cite{Jiang:2023nmd}, a criterion consists of  the coefficients of the free energy is used to determine the coexisting region in the phase diagram. It is clear that in the thermodynamic limit, the Taylor expansion coefficients of the free energy is related to the experimental cumulant ratios~\cite{Stephanov:2008qz,Mukherjee:2015swa}, thus in this article, we improve the criterion in \cite{Jiang:2023nmd} by connecting it with the experimental observables, namely, the cumulants.
%showed that the first-order phase transition can be written as a combination of the cumulants ratios, thus the value of criterion is possible to be determined by the experiment measurements. %As the cumulants are related to the Taylor coefficients of the effective potential, $\Delta$ in this criterion is written as a combination of the cumulants, and can be utilized to determine whether the first-order phase transition occurs or not by its sign.
In the thermodynamic limit, the criterion agrees with a recent study \cite{Lu:2025qyf}, where its sign determines the presence of coexisting phases in the system.
%A primary advantage of the criterion is that one needs not to analyze the collision energy-dependent behaviors like that for the cumulants, but rather determine the occurrence of the first-order phase transition barely based on its sign at every single collision energy.

Considering the realistic heavy-ion collision, the dynamical effects and finite-size effects on the final-state observables like the cumulants are significant based on the former studies~\cite{Berdnikov:1999ph,Stephanov:2009ra,Mukherjee:2015swa,Jiang:2017mji,Jiang:2017fas,Jiang:2021zla,Sakaida:2017rtj,Nahrgang:2018afz,Jiang:2023nmd,Pihan:2022xcl,brezin1985npb,Imry1980prb,Spieles:1997ab,Zabrodin:1998vt}.
Setting the dynamic system to be finite and spatially homogeneous, we study these effects on the criterion, and if the first-order phase transition signal is preserved during the dynamical evolution.
%Note that we assume the effective potential is supposed to depend only on the temperature and baryon chemical potential in the current stage. For simplicity, we also assume the temperature or baryon chemical potential distribution to be spatially homogeneous, the system volume is set to be constant or expanding in the dynamical evolution. More complicated simulations of the realistic inhomogeneous system (QGP fireball) will be left for future study.
The rest of the paper is organized as follows.
In Sec.\,\ref{sec2}, based on the parameterized Landau free energy that exhibits different phase transitions in the $\mu$-$T$ plane, we present the reduction of the cumulants-dependent criterion $\Delta$ for the first-order phase transition and the dynamical framework for the above task.
The dynamical Landau free energy and the dynamical cumulants are evaluated with the Fokker-Planck equation, from which we extract the non-equilibrium $\Delta$.
In Sec.\,\ref{sec3}, by setting the trajectories of the evolution in the $\mu$-$T$ plane, we illustrate how the different contributory factors influence the behaviors of $\Delta$ during the dynamical evolution process, including different phase transition scenarios, initial temperatures, volume size, relaxation rate, type of trajectories (curved or straight trajectories in the $\mu$-$T$ plane), and expanding volume. Then we show results of $\Delta$ on the hypothetical freezeout line.
In Sec.\,\ref{sec4}, we summarize the main results and make further discussions.

\section{The formalism}\label{sec2}
\subsection{the first-order phase transition criterion}
The QCD equation of state in the large chemical potential region is currently still unclear as the theories have not delivered solid model- and parameter-independent calculations.
Thus, instead of model-dependent free energy, we adopt parametric free energy in this article.
As the criticality of the QCD phase transition falls into the same universality class as the 3-dimensional Ising model~\cite{Hohenberg:1977,Son:2004iv}, the QCD free energy density in the vicinity of CEP can be analytically expanded as a function of the QCD order parameter field $\sigma$.
For practicability and simplicity, at least to the power of four, the free energy density can describe the phase structure of crossover and first-order phase transition at the same time.
Generally, the  free energy density is written as \cite{Jiang:2023nmd}:
\begin{equation}
\Omega[\sigma]=\alpha _{1}(\mu,T)\sigma +\frac{\alpha _{2}(\mu,T)%
}{2}\sigma ^{2}+\frac{\alpha _{3}(\mu,T)}{3}\sigma ^{3}+\frac{\alpha
_{4}(\mu,T)}{4}\sigma ^{4}, \label{pot}
\end{equation}
The constant term is omitted because it does not influence the structure of the free energy density and the phase diagram.
A finite $\alpha _{1}\sigma $ term is introduced to handle the explicit chiral symmetry breaking of the quark masses~\cite{Petropoulos:1999}.
The cubic term, $\alpha _{3}\sigma^3$, emerges after the renormalization contributed by the high-momentum modes of the $\sigma$ field.
The coefficient of the quartic term, $\alpha_{4}$, is supposed to be positive to sustain the stability of the system.
Note that the temperature and baryon chemical potential are supposed to be spatially homogeneous, and a zero-momentum mode approximation of the $\sigma$ field is assumed throughout the paper.

\begin{figure}[]
\centering
\includegraphics[width=0.95\columnwidth]{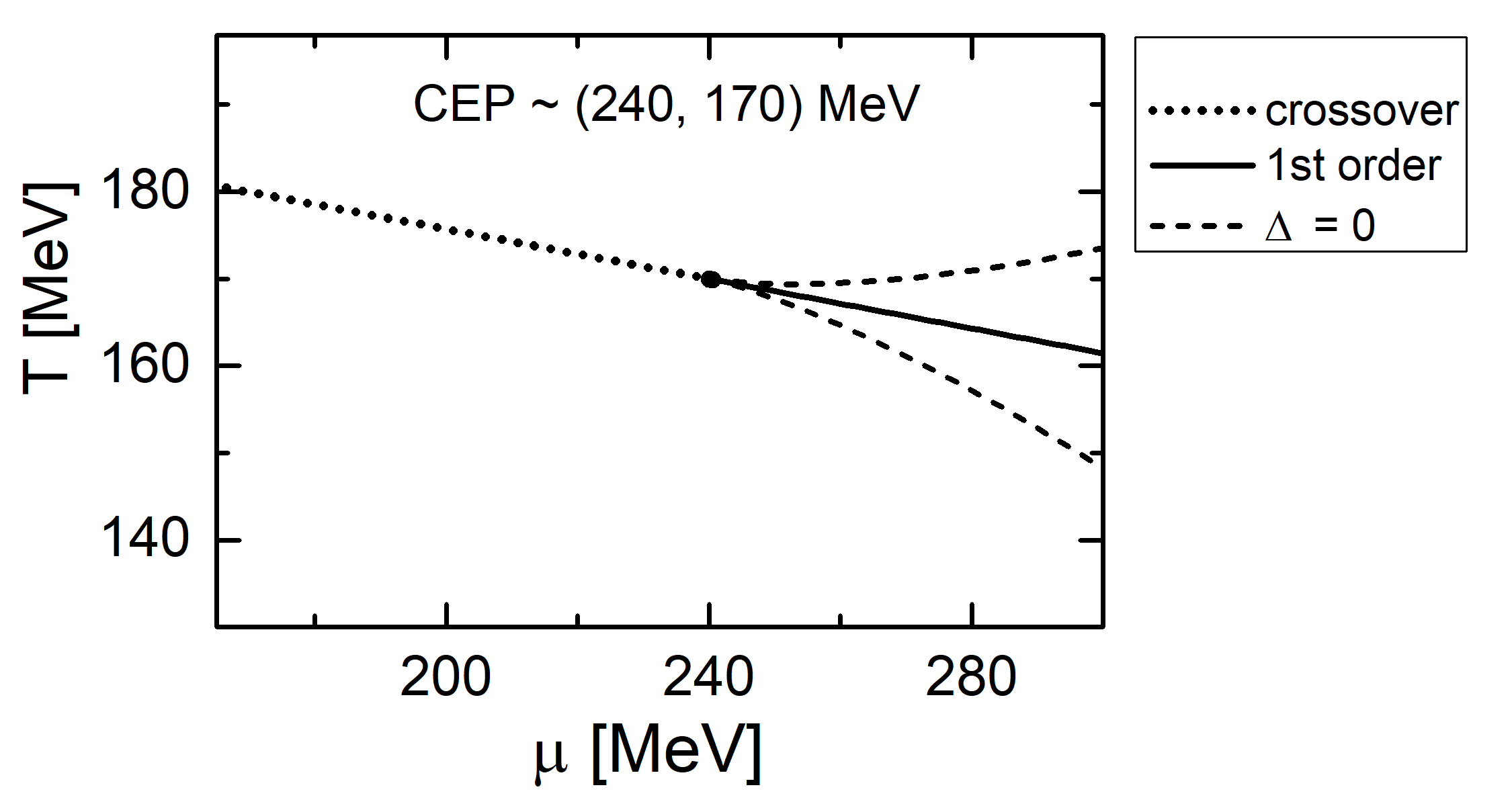}
\vspace*{-2mm}
\caption{The phase diagram calculated based on the parametric free energy in Eq.\,(\ref{pot2}). The dotted line represents the crossover line, the solid black line represents the first-order phase transition line, and the black dot in between is the CEP with its location at $(T_{c}, \mu_{c}) = (170, 240)\;$MeV. The black dashed lines, determined by $\Delta=0$ in the thermodynamic limit, mark the boundaries of the two-phase coexistence region.}
\label{phasediagram}
\end{figure}

Following the Ising parameterization method developed in Ref.~\cite{Jiang:2023nmd}, one may take a new variable $\tilde{\sigma} =\sigma-\sigma _{c}$ with $\sigma_{c}(\mu,T)=-{\alpha _{3}(\mu,T)}/(3\alpha _{4}(\mu,T))$ to the above free energy density to cancel the cubic term, it is then converted  to the well-known Ising form
\begin{equation}
\Omega \lbrack \tilde{\sigma}]=\eta _{1}^{} \left( T, \mu \right) \tilde{\sigma}+ \frac{1}{2}\eta _{2}^{} \left( T, \mu \right) \tilde{\sigma}^{2} + \frac{1}{4} \eta _{4}^{} \left( T, \mu \right) \tilde{\sigma}^{4}.  \label{pot2}
\end{equation}
The coefficients $\eta_{1}^{} \left( T, \mu \right)$ and $\eta_{2}^{} \left( T, \mu \right)$ is linearly parameterized in the $\mu$-$T$ plane as
\begin{eqnarray}
\eta_{1}^{} (T, \mu)& = & d_{1}[\left( \mu -\mu _{c}\right)\sin{\theta_{b}}
- \left( T - T_{c}\right)\cos{\theta_{b}} ] , \label{eta1d1}\\
\eta_{2}^{} (T, \mu)& = & d_{2}[ -\left( \mu - \mu _{c}\right)\sin{\theta_{b'}}
+\left( T - T_{c}\right)\cos{\theta_{b'} } ] ,\label{eta2d2}
\end{eqnarray}
Within reasonable ranges, the parameter set is again chosen as $(T_{c}, \, \mu _{c})=(170,\, 240)$ {MeV}, $\sigma_{c}=50 ~\rm{MeV}$, $d_{1}=3\times 10^{4} ~\rm{MeV}^{2}$, $d_{2}=400 ~\rm{MeV}$, $\eta_{4}=15$,    $\sin\theta_{b'} = -\cos\theta_b = 0.99$, and $\cos\theta_{b'} = \sin\theta_b= 0.141$ \cite{Jiang:2023nmd}. The coefficients $\alpha_i$ near the CEP are parameterized because of the following relations with $\eta_{i}$:
\begin{eqnarray}
\alpha _{1} & = & \eta _{1}^{} - \eta _{2}^{} \sigma _{c} - \eta_{4}^{} \sigma _{c}^{3}, \label{eta1}  \\
\alpha _{2} & = & \eta _{2}^{} + 3\eta _{4}^{} \sigma _{c}^{2},  \label{eta2} \\
\alpha _{3} & = & -3 \eta _{4}^{} \sigma _{c},  \label{eta3} \\
\alpha _{4} & = & \eta _{4}^{} .  \label{eta4}
\end{eqnarray}
The phase diagram based on the above parameter setup is plotted in  Fig.\,\ref{phasediagram},  with both the crossover (black dotted line) and the first-order phase transition (black solid line) included.
%{\color{red}\sout{As the coefficients of the free energy density are exactly accessible, mathematically, a criterion is available to specify the two-phase coexisting by combining these coefficients. Denote the criterion as $\Delta'$, it can be written as}
%\begin{equation}
%\Delta^{\prime} \equiv 27\eta_{1}^2 \eta_{4}^{2} + 4\eta_{2}^{3}\eta_{4}\, . \label{deltaprime}
%\end{equation}
%\sout{$\Delta^{\prime} <0$ corresponds to three unequal extreme values in the free energy density. With this criterion, we can separates a two-phases coexisting region in the $\mu-T$ plane, the boundaries of which ($\Delta^{\prime} = 0$) are marked by the black dashed lines in Fig.\,\ref{phasediagram}.}}

As the cumulants and cumulant ratios of $\sigma$ field play crucial roles in the search for QCD phase transition signal, in the thermodynamic limit, one can
expand $\Omega[\Tilde{\sigma}]$ around the global minimum $\tilde{\sigma}_0$\footnote{The global minimum $\tilde{\sigma}_0$ of the free energy density is determined by $\left. \frac{d\Omega \left[ \tilde{\sigma}\right] }{d\tilde{\sigma}}%
\right\vert _{\tilde{\sigma}=\tilde{\sigma}_{0}}=0$, in the first-order phase transition case, pick the one with lower  energy.}, and  rewritten it as \cite{Stephanov:2008qz,Mukherjee:2015swa}
\begin{equation}
    \Omega \left[ \tilde{\sigma}\right] =\frac{1}{2}m^{2}\left( \tilde{\sigma}%
-\tilde{\sigma}_0\right) ^{2}+\frac{1}{3}\lambda _{3}\left( \tilde{\sigma}-\tilde{\sigma}_0\right) ^{3}+\frac{1}{4}\lambda _{4}\left( \tilde{\sigma}-\tilde{\sigma}_0\right) ^{4}, \label{pot3}
\end{equation}
Comparing  $\Omega \left[ \tilde{\sigma}\right]$ with the free energy density in Eq.\,(\ref{pot2}), the relations between $(m, \lambda_3, \lambda_4)$ and $(\eta_1, \eta_2, \eta_4, \sigma_0)$ are derived,
\begin{eqnarray}
\eta _{1} &=&-\frac{m^{2}\lambda _{3}}{3\lambda _{4}}+\frac{2\lambda _{3}^{3}%
}{27\lambda _{4}^{2}},~~~~~\eta _{2}=m^{2}-\frac{\lambda _{3}^{2}}{3\lambda _{4}},
\nonumber \\
\eta _{4} &=&\lambda _{4},~~~~~~\sigma _{0}=\frac{\lambda _{3}}{3\lambda _{4}}.
\end{eqnarray}%
For a large volume, the cumulants $C_n (n\leq 4)$ can be expressed as functions of $m, \lambda_3,$ and $\lambda_4$ \cite{Stephanov:2008qz,Mukherjee:2015swa}
\begin{equation}
C_{2}=\frac{1}{V_{4}}\frac{1}{m^{2}}, ~~~
C_{3} =-\frac{2\lambda _{3}}{V_{4}^{2}}\frac{1}{m^{6}},  ~~~C_{4}=\frac{6}{%
V_{4}^{3}}\left( \frac{2\lambda _{3}^{2}}{m^{2}}-\lambda _{4}\right) \frac{1%
}{m^{8}},  \label{cumulants}
\end{equation}
where $V_4=V/T$.

From the free energy density shown in Eq.\,(\ref{pot3}), we can judge whether a coexistence phase region exists by counting the number of real roots of the following condition for extrema:
\begin{equation}
\frac{\partial{\Omega \left[ \tilde{\sigma}\right]}}{\partial{\tilde{\sigma}}}=m^{2}\left( \tilde{\sigma}-\tilde{\sigma}_0\right) +\lambda _{3}\left( \tilde{\sigma}-\tilde{\sigma}_0\right) ^{2}+\lambda _{4}\left( \tilde{\sigma}-\tilde{\sigma}_0\right) ^{3}=0.
\end{equation}
There is always one root located at $\tilde{\sigma}_0$, while the existence of other real roots is determined by the condition $\lambda_3^2-4\lambda_4 m^2 >0$. Using the relations in Eq.\,\eqref{cumulants}, we can re-express the criterion for the first-order phase transition in a neat form as function of cumulant ratios:
\begin{equation}
\Delta =\frac{ 8C_{4}/C_{2}-21 (C_{3}/C_{2})^{2} }{12(C_{2}/C_{1})^{4}}>0. \label{delta}
\end{equation}
The resulted $\Delta$ depends on the system volume. As has been discussed in Ref.~\cite{Jiang:2023nmd}, when the size of the system is larger than $10^{5}\,\textrm{fm}^{3}$, the calculated cumulants are basically in accord with the results \eqref{cumulants} in the perturbative expansion around a Gaussian distribution~\cite{Stephanov:2008qz}. However, when the size is small, the finite volume of the system has great influence on the behaviors of the cumulants. In this case, the cumulants should be obtained from the probability distribution function $P[\sigma]\sim\exp[-V_4 \Omega[\sigma]]$ directly. Statistically, $\sigma$'s moments $\mu_n$ are derived using
\begin{eqnarray}
\mu_{n}  = \int \textrm{d} \sigma \sigma^{n} P[\sigma] {\Bigg/} \int \textrm{d}\sigma  P[\sigma] \,
\label{eq:ob}\end{eqnarray}
and the cumulants are calculated through the following relations:
\begin{eqnarray}
  C_{1} &=& \mu_{1}, \\
  C_{2} &=& \mu_{2}-{\mu_{1}}^{2}, \\
  C_{3} &=& \mu_{3}-3\mu_{2}\mu_{1}+2{\mu_{1}}^{3}, \\
  C_{4} &=& \mu_{4}-4\mu_{3}\mu_{1}-3{\mu_{2}}^{2}+12\mu_{2}{\mu_{1}}^{2}-6{\mu_{1}}^{4}.
\end{eqnarray}
As we will show in the following, a small volume will significantly affect the criterion $\Delta$.

\subsection{dynamical approach}

%{\it{dynamical approach:}}
As demonstrated in the early studies~\cite{Berdnikov:1999ph,Stephanov:2009ra,Mukherjee:2015swa,Jiang:2017mji,Jiang:2017fas,Jiang:2021zla,Jiang:2023nmd}, the dynamical effects and finite-size effects have great influences on the dynamical cumulants in the phase transition region.
In principle, the dynamical free energy can be derived from the real-time formalism of the finite temperature path integral, based on which we can discuss these effects on the cumulants and $\Delta$, but the deduction is relatively challenging from the theoretical side.
In the following, we choose the alternative and more effective Fokker-Planck equation to derive the dynamical free energy density, make calculations on $\Delta$, and further explore kinds of effects on the criterion.
The time evolution of the probability distribution $P[\sigma;t]$ for the $\sigma$ field reads~\cite{Mukherjee:2015swa,Jiang:2023nmd},
\begin{equation}\label{evolution}
\partial_{t} P[\sigma;t] = - \frac{1}{m_{\sigma}^{2} \tau_{\rm{eff}}}
\partial_{\sigma} \left\{\partial_{\sigma}\left(\Omega[\sigma;t] -
\Omega[\sigma] \right) P[\sigma;t] \right\},
\end{equation}
where $\Omega[\sigma;t]\equiv -\frac{T}{V} {\rm ln} P[\sigma;t]$ is the dynamical free energy density, and $\Omega \left[\sigma \right]$ is the instantaneous equilibrium free energy density as shown in Eq.\,\eqref{pot} at specific $T$ and $\mu$. $m_{\sigma}$ denotes the equilibrium mass of the $\sigma$ field which is defined as $m_{\sigma}^{2} =\frac{\mathrm{d}^{2} \Omega[\sigma]}{\mathrm{d}\sigma^{2}}{\big |}_{\sigma =\sigma_{0}} \, ,$
with $\sigma_{0}$ being the global minimum of $\Omega[\sigma]$. $\tau_{\rm{eff}}$ is the effective relaxation rate,
and the dependence of the relaxation rate $\tau_{\rm{eff}}$ on the equilibrium correlation length $\xi_{\rm{eq}}(T,\mu)$ is assumed to satisfy $\tau_{\rm{eff}} = \tau_{\rm{rel}}\left(\frac{\xi_{\rm{eq}}}{\xi_{\rm{ini}}}\right)^{z} \, ,$
where $\xi_{\rm{eq}} = m_{\sigma}^{-1}$~\cite{Mukherjee:2015swa}.
$\tau_{\rm{rel}}$ and $\xi_{\rm{ini}}$ are the initial relaxation rate and the initial equilibrium correlation length away from the phase transition region. The initial relaxation rate is a free parameter that is input by hand in the later calculations.
The value $z=3$ is given by the dynamical critical exponent of Model H~\cite{Hohenberg:1977,Son:2004iv}.
As the general form of the free energy density is assumed to be not affected by the dynamical evolution and the size of the system, one may also expand the dynamical free energy density as $\Omega[\sigma;t]= \sum_{i} \zeta_{i}(t)\sigma^{i} /{i}$. Retaining the same order as its counterpart, the dynamical free energy density is also truncated at the fourth order. A general form of the dynamical equation for $\zeta_i(t)$ is derived in the earlier work \cite{Jiang:2023nmd}. Explicitly,
the time evolution of  $\Omega[\sigma;t]$'s coefficients  $\zeta_{j} (j=1,\cdots,4)$  are given by
\begin{eqnarray}
\frac{\mathrm{d} \zeta_{1}(t)}{\mathrm{d} t} & = &\zeta_{1}(t)   \label{eq12}
\partial_{t} \left(\rm{ln}\frac{T}{V}\right) + \frac{2 T}{V} \frac{[ \zeta_{3}(t)- \alpha_{3}]}{m_{\sigma}^{2} \tau_{\rm{eff}}} \nonumber \\
&& -\frac{ \zeta_{2}(t) [\zeta_{1}(t)- \alpha_{1} ]}{m_{\sigma}^{2} \tau_{\rm{eff}}}-\frac{ \zeta_{1}(t) [\zeta_{2}(t) - \alpha_{2}]}{m_{\sigma}^{2} \tau_{\rm{eff}}},\\
\frac{\mathrm{d}\zeta_{2}(t)}{\mathrm{d} t} & = &\zeta_{2} (t)
\partial_{t}\left(\rm{ln}\frac{T}{V}\right)+ \frac{6 T}{V} \frac{[ \zeta_{4}(t)- \alpha_{4}]}{m_{\sigma}^{2} \tau_{\rm{eff}}} \nonumber \\
&& -\frac{ 2\zeta_{3}(t) [\zeta_{1}(t)- \alpha_{1} ]}{m_{\sigma}^{2} \tau_{\rm{eff}}}
-\frac{ 2\zeta_{2}(t) [\zeta_{2}(t) - \alpha_{2}]}{m_{\sigma}^{2} \tau_{\rm{eff}}}\nonumber \\
&&
-\frac{ 2\zeta_{1}(t) [\zeta_{3}(t) - \alpha_{3}
]}{m_\sigma^2 \tau_{\rm{eff}}},\\[1mm]
\frac{\mathrm{d}\zeta_3(t)}{\mathrm{d} t} & = &\zeta_{3}(t)
\partial_{t} \left(\rm{ln}\frac{T}{V}\right) - \frac{ 3\zeta_{4}(t) [\zeta_{1}(t)- \alpha_{1}]}{m_{\sigma}^{2} \tau_{\rm{eff}}}\nonumber \\
&&  -\frac{ 3\zeta_{3}(t) [\zeta_{2}(t) - \alpha_{2} ]}{m_{\sigma}^{2} \tau_{\rm{eff}}}
-\frac{ 3\zeta_{2}(t) [\zeta_{3}(t) - \alpha_{3} ]}{m_{\sigma}^{2} \tau_{\rm{eff}}}\nonumber \\
&& -\frac{ 3\zeta_{1}(t) [\zeta_{4}(t) - \alpha_{4} ]}{m_{\sigma}^{2} \tau_{\rm{eff}}},\\[1mm]
\frac{\mathrm{d}\zeta_4(t)}{\mathrm{d} t} & = &\zeta_4(t)
\partial_{t} \left(\rm{ln}\frac{T}{V}\right) - \frac{ 4\zeta_{4}(t) [\zeta_{2}(t)- \alpha_{2} ]}{m_{\sigma}^{2} \tau_{\rm{eff}}}\nonumber \\
&&   -\frac{ 4\zeta_{3}(t) [\zeta_{3}(t) - \alpha_{3} ]}{m_{\sigma}^{2} \tau_{\rm{eff}}}
-\frac{ 4\zeta_{2}(t) [\zeta_{4}(t) - \alpha_{4} ]}{m_{\sigma}^{2} \tau_{\rm{eff}}}. \label{eq15}
\end{eqnarray}

By setting up the parameters and the evolving trajectories, we can numerically solve the coupled equations Eqs.\,(\ref{eq12})-(\ref{eq15}) and obtain the dynamical free energy density, and further explore the behaviors of equilibrium or dynamical $\Delta$ according to Eq.\,(\ref{delta}).
Note that the cumulants defined above are associated with the order parameter $\sigma$, as we start from the free energy and probability distribution function for the $\sigma$ field, however, a similar procedure can be developed if one parameterizes the free energy with respect to other order parameters, such as the number density.

\section{Numerical results}
\label{sec3}

In the above section, we have described the theoretical framework for the calculation of dynamical and equilibrium $\Delta$, as well as the setting up of the parameters. In the following, we will present the numerical results of $\Delta$ along specific trajectories for the dynamical evolution. For most of the discussions, we adopt a vertical evolution trajectory in the $\mu$-$T$ plane,
the baryon chemical potential of which is a fixed numeric value, and the temperature decreases as a function of time in the following way \cite{Mukherjee:2015swa}
\begin{equation}\label{temp}
T(t)=T_{0} \left( \frac{t + t _{0}}{t _{0}}\right)^{-\lambda},
\end{equation}
where $T_{0}$ is the initial temperature for the evolution, $t _{0} = 10\,\rm{fm}$ is the initial time, and the exponent is set to be $\lambda = 0.45$~\cite{Mukherjee:2015swa}. We will also present results with curved trajectories later, in which the chemical potential $\mu$ is temperature-dependent (i.e. time-dependent).
For the choice of system volume, in the first four subsections, we set it to be finite and the magnitude is (or close to) $10^3$ fm$^3$ (the typical size of the QGP fireball at RHIC\,\cite{Song:2007ux}); in the last two subsections and the appendix, we set the volume to expand over time. Note again that the spatial inhomogeneity is beyond the scope of our discussion in the current stage.

\begin{figure}[t]
\centering
\includegraphics[width=0.86\columnwidth]{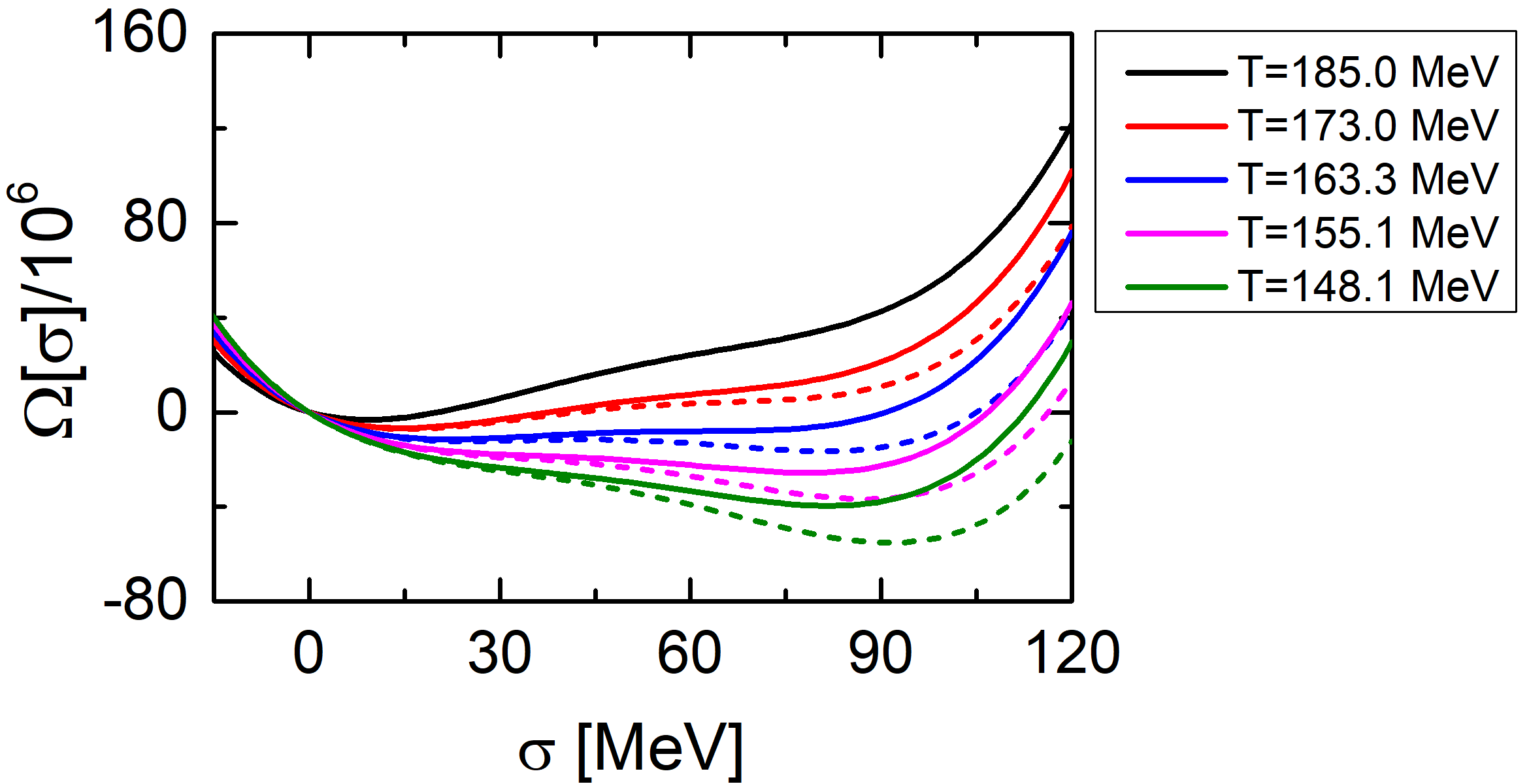}
\vspace*{-2mm}
\caption{(color online) Calculated equilibrium free energy density $\Omega[\sigma]$ (the dashed lines) and the dynamical $\Omega[\sigma,t]$ (the solid lines) as a functions of $\sigma$ at several temperatures, with the baryon chemical potential $\mu = 270 $ MeV and $\tau_{rel} = 0.1$ fm.  The initial temperature for the dynamical evolution is at $T=185$ MeV .}
\label{potential}
\end{figure}

Before the calculation of $\Delta$, we make comparisons of dynamical and equilibrium free energy density. By setting the baryon chemical potential $\mu = 270$ MeV ($> \mu_c$), the initial relaxation rate $\tau_{rel} = 0.1$ fm, the system volume $V = 10^3$ fm$^3$, and the initial temperature $T_0=185$ MeV which is higher than the phase transition temperature, the system starts evolving, undergoes a first-order phase transition and ceases evolving at a low temperature. The equilibrium  $\Omega[\sigma]$ (the dashed lines) and dynamical $\Omega[\sigma,t]$  (the solid lines) as functions of $\sigma$ at several temperatures are shown in Fig.\,\ref{potential}. The typical two-phase coexisting structure appears for both the equilibrium and the dynamical case. The shape of the free energy density slightly changes during the evolution because of the cutoff in Eqs.\,(\ref{eq12})-(\ref{eq15}), but generally, the dynamical free energy density follows the steps of the equilibrium one as the system passes through the phase transition region.

\begin{figure*}[t]
\centering
\includegraphics[width=2.\columnwidth]{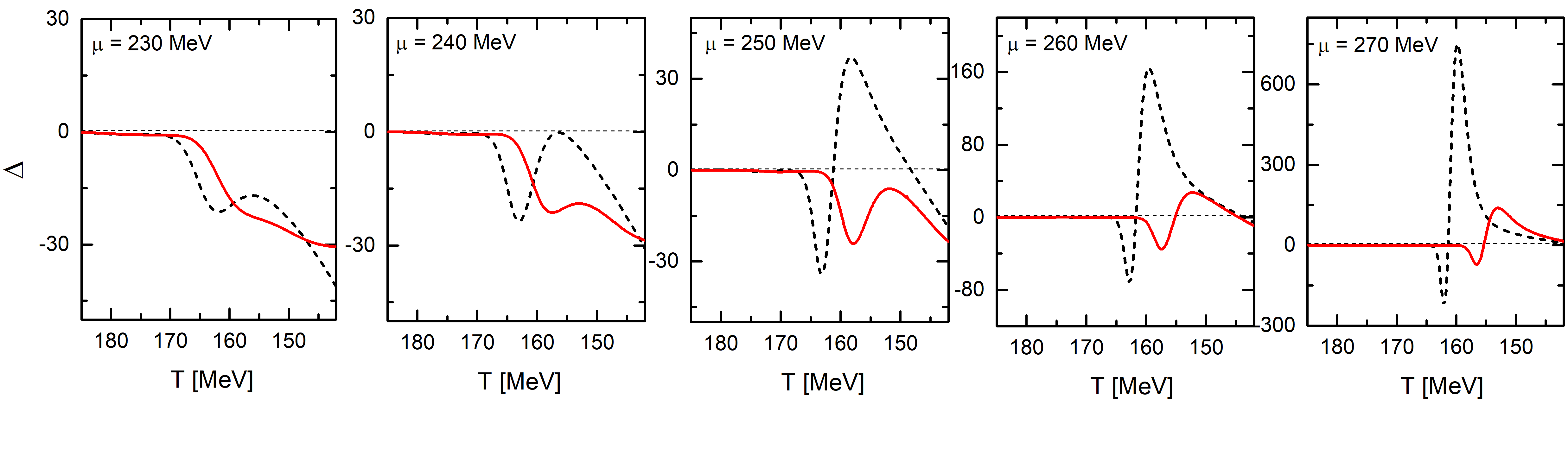}
\vspace*{-2mm}
\caption{(color online) The equilibrium  and  dynamical $\Delta$ as functions of the temperature $T$, with the system volume set at $V = 10^{3}\,\textrm{fm}^{3}$ and the initial relaxation rate $\tau_{rel} = 0.1\;$fm. The dashed lines represent the instantaneous equilibrium results and the red solid lines represent the dynamical results. In different subfigures, the evolution trajectories are set to be along different constant $\mu$ lines in the $\mu$-$T$ plane. }
\label{delta-mu}
\end{figure*}
\subsection{different phase transition scenarios}
From this subsection, we study several realistic factors' influence on the criterion. First, with the setup of $V = 10^3$ fm$^3$, $\tau_{rel}=0.1$ fm/c and the temperature $T(t)$ evolves as that in Eq.\,(\ref{temp}), we select several different baryon chemical potentials on both sides of the CEP so that the system evolves along different vertical trajectories and crosses the crossover, CEP or first-order phase transition line, respectively.

The numerical results of $\Delta$ as functions of the temperature are presented in Fig.\,\ref{delta-mu}, where the dotted lines and the solid lines represent the instantaneous equilibrium (corresponding to relaxation time $\tau_{ref} \rightarrow 0$) and dynamical results, separately.
Of great interest is the temperature interval with positive $\Delta$, which marks the appearance of coexisting phases. %Note that in the equilibrium case, the positive $\Delta$ region does not overlap the negative $\Delta^{\prime}$ region as a result of the finite-size effects as mentioned above.
%Note that the positive $\Delta$ region appears at a lower temperature because the lower temperature effectively discounts the finite-size effects through the distribution $\exp[-V \Omega[\sigma;t]/T]$.
Due to the memory effects, the dynamical $\Delta$ follows and reproduces similar behaviors as the corresponding equilibrium ones. But the positive $\Delta$ region are lagged and appears at a lower temperature because of a finite relaxation time.
Besides, the temperature interval for positive $\Delta$ shrinks and is depressed during the dynamical evolution. Particularly, if the period of coexistence of two (equilibrium) phases is only a small part of the evolution process, the positive $\Delta$ signals disappear during the evolution, as shown in the subfigure with $\mu = 250$ MeV. For larger chemical potential cases, with a long enough coexisting period during the evolution, the signal survives and appears in a lower-temperature region, as shown in the subfigure with $\mu = 260\;$MeV and $\mu = 270\;$MeV.

Note that the final state (referring to the state where the system chemically freezes out) is observable in experiments, the temperature of which is below the phase transition temperature. Based on the results in Fig.\,\ref{delta-mu}, the combination of finite-size effects and dynamical effects on $\Delta$ leads to the appearance of positive $\Delta$ signals at a temperature lower than the phase transition one, meaning they are likely to be detected for some appropriately chosen trajectories.
%Generally, from the above numerical results of $\Delta$, we find that the two quantities exhibit similar behaviors during the dynamical evolution. But from the point of view of the experimental side, $\Delta$ is more favorable for experimental observation at RHIC, as it is related to the cumulant ratios.  Thus in the following subsections, our discussions are specifically focused on the behaviors of $\Delta$.

\subsection{different volume and relaxation rate} \label{3.2}

\begin{figure}[ptb]
\centering
\includegraphics[width=0.9\columnwidth]{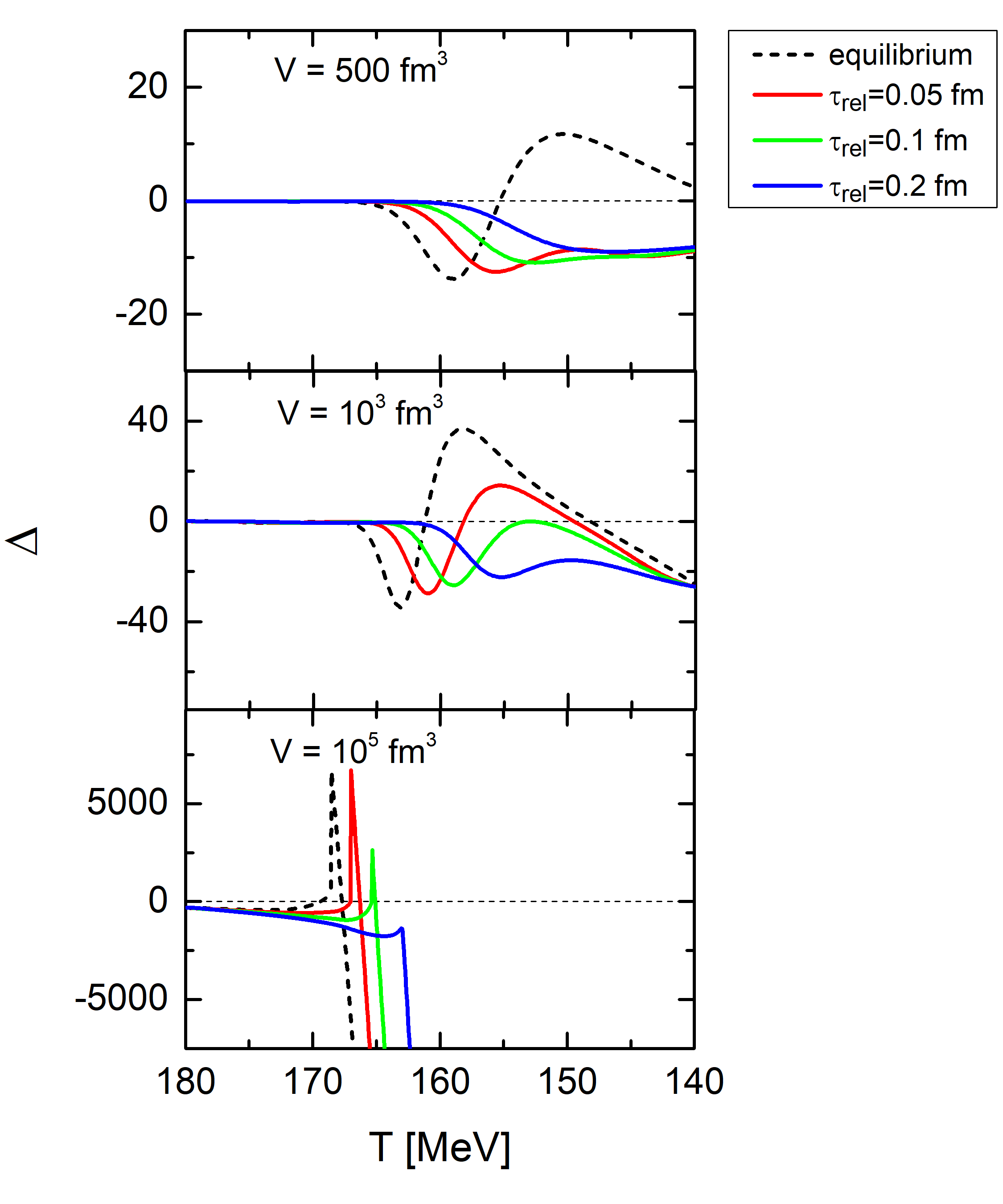}
\vspace*{-4mm}
\caption{(color online) The calculated dynamical results of $\Delta$ (the colored lines) and the equilibrium results (the dashed lines) as functions of the temperature with the baryon chemical potential $\mu=250\,$MeV. The different colored lines are for the results with initial relaxation rate $\tau_{rel} = 0.05, 0.1, 0.2\,$fm, respectively. From the upper to the lower panel, the results are for the system with volume $V = 500, 10^{3}, 10^{5}\,\textrm{fm}^{3}$, respectively.  }
\label{delta-Vtau}
\end{figure}

As mentioned above, the size of the system has a great influence on the behaviors of $\Delta$. At the same time, the relaxation rate is also a key factor that affects the speed of the dynamical evolution. In this subsection, we combine the two factors and discuss their effects on $\Delta$. Fig.\,\ref{delta-Vtau} visually illustrates how $\Delta$ is influenced by these factors. The evolution trajectory is set to be along the $\mu = 250\,$MeV line. From the upper to the lower panel, $\Delta$ is calculated as a function of temperature, and the volume is $V = 500$, $10^{3}$ and $10^{5}\,\textrm{fm}^{3}$, separately.
In each subfigure, the dashed line represents the equilibrium $\Delta$, the red, green, and blue lines are the dynamical $\Delta$ with initial relaxation rate $\tau_{rel} = 0.05, 0.1, 0.2\,$fm, respectively.

For the equilibrium $\Delta$ (dashed lines) in different subfigures, the positive $\Delta$ region shifts to a lower temperature due to a finite size effect. At the same time, the temperature interval (fixed $\mu$) for the positive $\Delta$ widens as the system size $V$ decreases. We analyze  these finite-size effects below. Note that $\Delta$, as a function of cumulants $C_n$, is fully determined by the volume-dependent probability distribution $ P[\sigma] \propto \exp[-V \Omega[\sigma]/T] $.  Focusing on the maximum of $\Delta$, a minor decrease in volume $ V $ can be partially compensated by decreasing the temperature to preserve $(V - dV) \Omega[\sigma] / (T - dT)$. As a result, the peak of $\Delta$ shifts toward lower temperatures as $V$ decreases. However, since $\Omega[\sigma]$ also depends on $ T $, view the subplots from bottom to top, the value of $V/T$ at the maximum of $\Delta$ remains decrease with decreasing $ V $. This leads to, as demonstrated in Ref.\,\cite{Jiang:2023nmd}, the broadening of the distribution $P[\sigma]$, as well as  $ C_n $ values.  Consequently, the temperature interval for positive $\Delta $ widens as $V$ decreases.

For all the dynamical $\Delta$ shown in the three subfigures, there are clear memory effects during the dynamical evolution, where the dynamical $\Delta$ follows the steps of the equilibrium ones and reproduces similar structures at a later time. Interestingly, for the current choices of the initial relaxation rate $\tau_{rel}$, the positive $\Delta$ signal disappears in the $V = 500$ fm$^3$ subfigure and is reserved for the other two volume cases when $\tau_{rel}$ is not too large ($\tau_{rel}=0.05$ and $0.1$ fm). In addition, the positive $\Delta$ signal is suppressed as the increase of $\tau_{rel}$, and when $\tau_{rel}=0.2$ fm, the positive $\Delta$ signal is faded away during the relaxation of the system for all the three cases. In short, both the decrease of the system size and the increase of the relaxation rate lead to the suppression of the positive $\Delta$ signal.

\subsection{different initial temperature}
\begin{figure}[tpb]
\centering
\includegraphics[width=0.96\columnwidth]{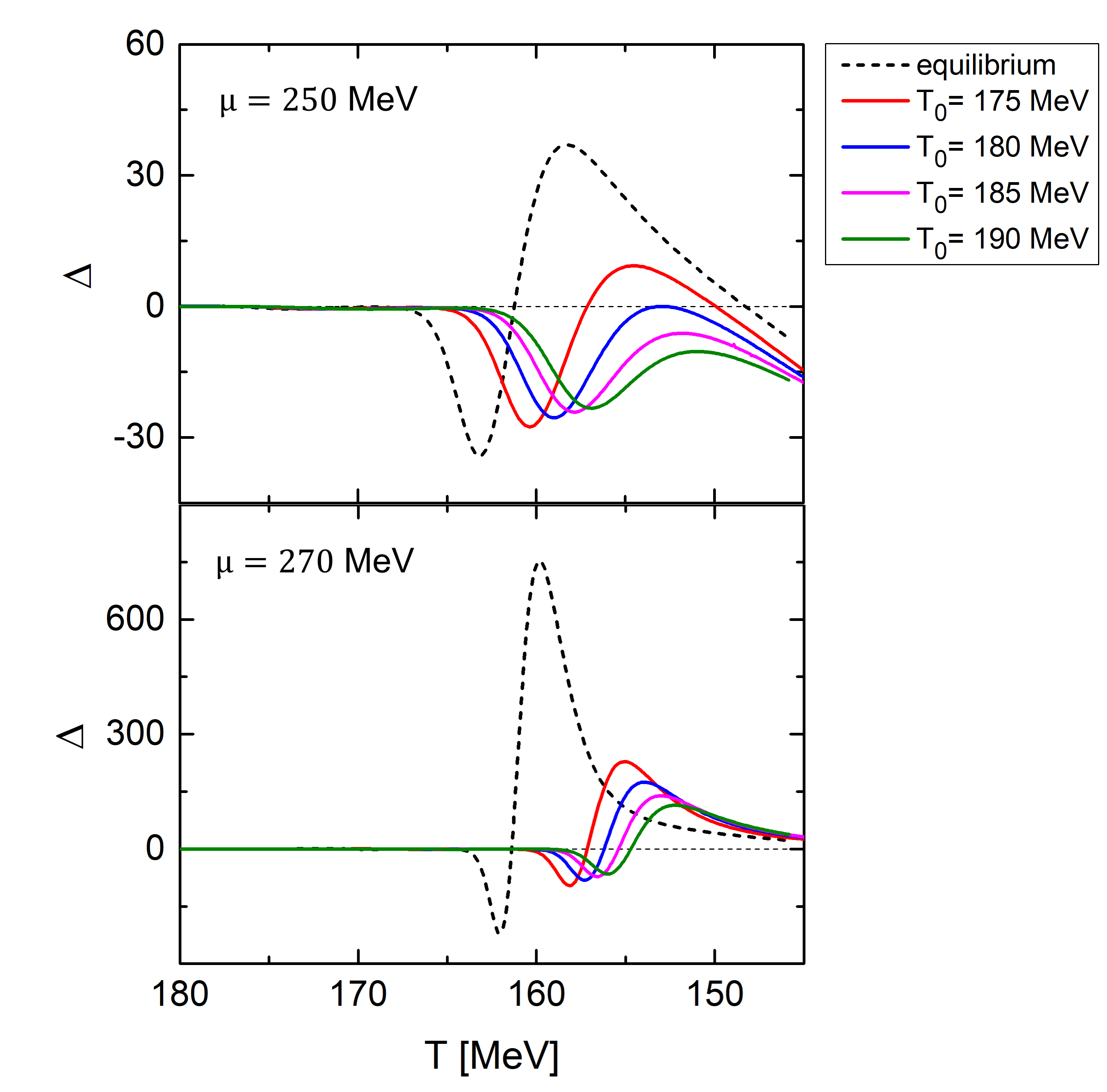}
\vspace*{-3mm}
\caption{(color online) The dynamical $\Delta$ (the colored lines) as functions of the temperature with initial temperature $T_{0}=175, 180, 185, 190\;$MeV respectively. The dashed line is the equilibrium result for comparison. The baryon chemical potential is set to be $\mu=250\;$MeV (upper panel) and $\mu=270\;$MeV (lower panel) with a volume of $V = 10^{3}\,\textrm{fm}^{3}$ and an initial relaxation rate of $\tau_{rel}=0.1\,$fm.   }
\label{delta-T0}
\end{figure}

In the realistic heavy-ion collision process, the initial temperature of the fireball evolution is determined by the collision energy, or, in other words, by how fast the two nuclei go across each other. At RHIC energies, the initial temperature in the center of the fireball has a difference of several tens of MeVs \cite{Song:2007ux}. Therefore, it is meaningful to investigate whether the initial temperature affects the behaviors of $\Delta$ during the dynamical evolution.

By setting the system volume to be $V = 10^{3}\,\textrm{fm}^{3}$ and the initial relaxation rate to be $\tau_{ref} = 0.1\,$fm, we calculate the dynamical $\Delta$ with initial temperature $T_{0} = 175, 180, 185, 190\,$MeV, respectively. The evolution trajectory is set to be along the $\mu = 250\,$MeV line. The results are illustrated in Fig.~\ref{delta-T0}, where the colored solid lines are the results of the dynamical $\Delta$, and the dashed line is the equilibrium $\Delta$ result for comparison.
As shown in Fig.\,\ref{delta-T0}, the dynamical $\Delta$ results neatly align along the transverse axis as the temperature decreases, demonstrating clear memory effects. For the case with $T_{0} = 175\,$MeV, which is closest to both the two-phase coexisting region and the phase transition line in the current setup, it takes a shorter time for the system to undergo the first-order phase transition, thus the positive $\Delta$ signal appears at a much earlier time (and higher temperature) than the others. At the same time, the temperature interval of positive $\Delta$ shrinks as the increase of the initial temperature and eventually disappears if the initial temperature is too far away from the phase transition region, as displayed by the lines of $T_{0}=185$ and $T_{0}=190$ MeV.

Note that along the evolution trajectory $T(t)$ [Eq.\,\eqref{temp}] for fixed $\mu$, the time interval spent in the mixed phase for $ T_0 = 190 $ MeV is indeed slightly longer than that for $ T_0 = 175 $ MeV. However, for $ T_0 = 175 $ MeV, the system rapidly enters the first-order phase transition region, allowing early recording and retention of the positive $\Delta$ signal. In contrast, when starting with a larger initial temperature, such as $ T_0 = 190 $ MeV, the dynamical evolution begins at a temperature well above the mixed-phase region. The initial value of $\Delta$ is negative. As the temperature decreases and passes through the phase transition region, the finite relaxation rate introduces a retarded response. Due to this delay, the system retains memory of its earlier state with negative $\Delta$ for an extended period. Even when passing through the region where the equilibrium $\Delta$ is positive, there is insufficient time for the dynamical $\Delta$ to transition from negative to positive. Consequently, the positive $\Delta$ signal (expected in equilibrium) is missing during the evolution for cases with high initial temperatures.
For comparison, in the lower panel ($\mu = 270$ MeV), where the chemical potential is larger, the first-order phase transition region expands. This prolongs the duration of the mixed-phase stage, enabling the positive $\Delta$
signal to appear even for the case with $ T_0 = 190 $ MeV.

\subsection{along curved trajectories}

\begin{figure}[t]
\centering
\includegraphics[width=0.8\columnwidth]{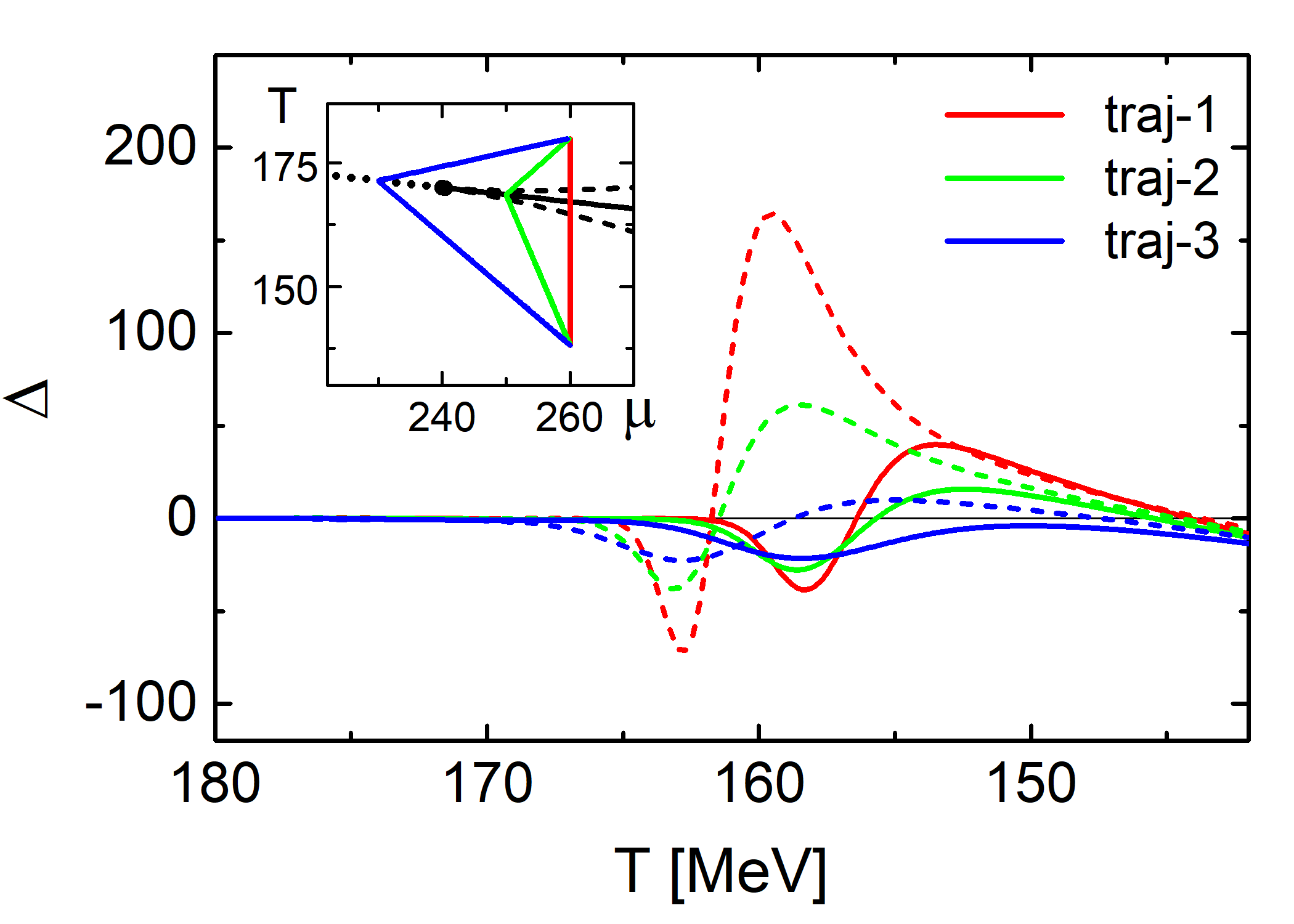}
\vspace*{-3mm}
\caption{(color online) The colored solid lines represent the dynamical  $\Delta$ along the three evolution trajectories in the $\mu$-$T$ plane shown in the inset, and the dashed lines are the equilibrium $\Delta$ results for comparison. The volume is set to be $V = 10^3$ fm$^3$, and the initial relaxation rate $\tau_{rel}=0.1$ fm. }
\label{delta-3trajs}
\end{figure}

In the above subsections, we adopt linear evolution trajectories with fixed baryon chemical potential, but it is known that in relativistic heavy-ion collisions, the system evolves along isentropic trajectories, which have the typical "$<$"-shape. For this reason, we study the $\Delta$ results along curved trajectories in this part. As the free energy density in Eq.\,(\ref{pot}) is too simplified to reproduce the typical isentropic trajectories like that in the effective theories, and on the other hand, the realistic isentropic trajectories, as well as the QCD equation of state, are not determined so far, we choose by hand the alternative "$<$"-shape trajectories and make discussions based on them.
The three different evolving trajectories are plotted in the inset of Fig.~\ref{delta-3trajs}. Traj-1 is set to be the vertical red line with $\mu =260\,$MeV, the other two curved trajectories are supposed to turn around at the phase transition line. The turning point for traj-2 is at $\mu =250\,$MeV, and that for traj-3 is at $\mu =230\,$MeV (the green and blue line). Note that the starting point and the ending point for the three trajectories are assumed to be the same. As the evolution of the system is controlled by the decreasing of temperature, the evolution time for all three trajectories is the same, which is $8$ fm.

Setting the system size to be $V = 10^{3}\,\textrm{fm}^{3}$ and the initial relaxation rate as $\tau_{rel}=0.1\;$fm, the decreasing of temperature is kept as in Eq.\,(\ref{temp}), and the time-dependent baryon chemical potential is obtained by parameterizing the three trajectories in the $\mu$-$T$ plane, we present the results of the equilibrium and dynamical $\Delta$ along the three trajectories in Fig.~\ref{delta-3trajs}.
The dashed lines represent the equilibrium $\Delta$ and the solid lines represent the dynamic results, with the red, green, and blue line corresponding to the result for traj-1, traj-2, traj-3, respectively.
Generally, the dynamical $\Delta$ results follow the trends of the equilibrium ones due to the memory effects, but the signals are reduced. Both the equilibrium and the dynamical $\Delta$ are suppressed when the trajectories are further away from the two-phases coexisting region. The positive dynamical $\Delta$ signal disappears for the traj-3 case, in which the evolving trajectory bypasses the CEP from the crossover side.

It is worth stressing that even for traj-3, which appears to lie entirely outside the first-order phase transition region, positive $\Delta$ signals still emerge at lower temperatures in the equilibrium case. The reason is the same as discussed in Sec.\,\ref{3.2}: due to finite-size effects, the temperature interval for positive $\Delta$ signals shifts to lower temperatures and broadens in a finite system.  The lower portion of traj-3 happens to fall within this shifted region where  $\Delta$ remains positive.

\subsection{with an expanding volume}

\begin{figure}[tb]
\centering
\includegraphics[width=0.8\columnwidth]{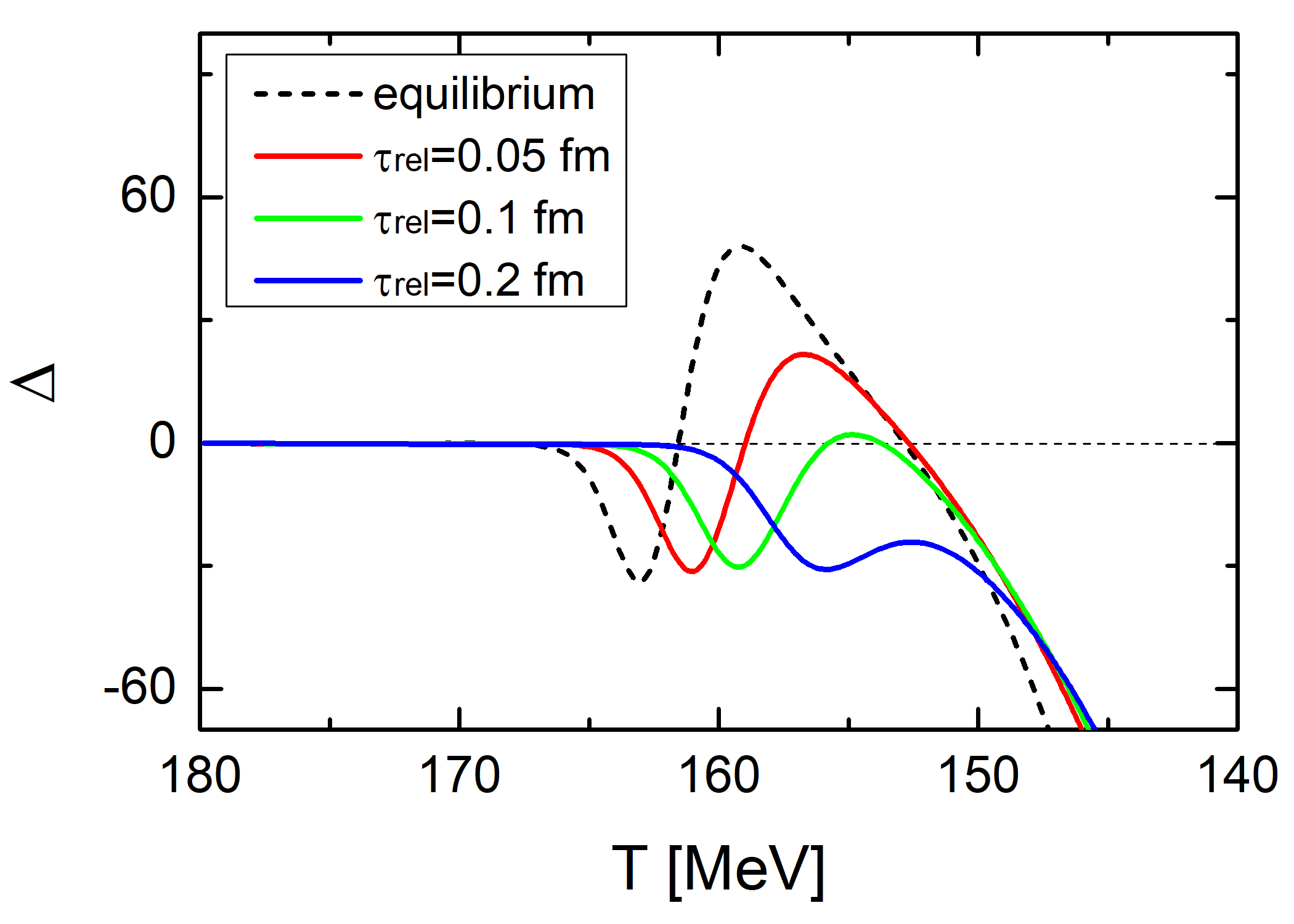}
\vspace*{-3mm}
\caption{(color online) The calculated dynamical results of $\Delta$ (the colored lines) and the equilibrium results (the dashed lines) as functions of the temperature with the baryon chemical potential $\mu=250\,$MeV. The different colored lines are for the results with initial relaxation rate $\tau_{rel} = 0.05, 0.1, 0.2\,$fm, respectively.}
\label{delta-tau-Vt}
\end{figure}

The volume of the system is set to be constant in the above discussions. To be closer to the experimental reality, hereafter, we make discussions on $\Delta$ by assuming the system is expanding, the increase of the system volume is Hubble-like\,\cite{Mukherjee:2015swa}, which has the form
\begin{equation}
    V(t)=V_{0} \left( \frac{t + t _{0}}{t _{0}}\right)^{3}. \label{volume}
\end{equation}
The initial volume of the system at $t_0$ is set to be $V_0=8^3 \,\rm{fm}^3$ and the final volume is approximately $14^3 \,\rm{fm}^3$ in the end of the evolution. The evolution trajectory is set to be along the $\mu=250$ MeV line. We present the numerical results of $\Delta$ as a function of temperature in Fig.\,\ref{delta-tau-Vt}.
Except the setup of volume, the other parameters are the same as those in Fig.\,\ref{delta-Vtau}.
The dashed line is the result of equilibrium $\Delta$, and the colored lines are the results of dynamical $\Delta$ with initial relaxation rate $\tau_{rel} = 0.05, 0.1, 0.2\,$fm, respectively (corresponds to the red, green and blue line).
As the magnitude of the system volume is around $10^3 \,\rm{fm}^3$ during the evolution, we compare these results with the $V=10^3 \,\rm{fm}^3$  subfigure in Fig.\,\ref{delta-Vtau}. For the equilibrium $\Delta$, the temperature interval for the positive $\Delta$ ends slightly earlier than that in Fig.\,\ref{delta-Vtau}, and the magnitude of $\Delta$ is larger, both of which are because of the increase of the volume size during the evolution. Except that, the behaviors of the dynamical $\Delta$ are qualitatively in accord with the results in Fig.\,\ref{delta-Vtau}.

With the setup of an expanding volume, we also make calculations on $\Delta$ for different phase transition scenarios (to compare with the results in Fig.\,\ref{delta-mu}); or with the trajectory fixed at $\mu = 250$ MeV, but  different initial temperatures (to compare with the results in Fig.\,\ref{delta-T0}); or along curved trajectories with the same starting and ending points (to compare with the results in Fig.\,\ref{delta-3trajs}). The numerical results are qualitatively in accord with the former discussions,  only with small changes in the temperature interval. One can refer to the appendix for more details of these results.

\subsection{on the hypothetical freeze-out line}

As the phase transition signals are usually observed after the system freezes out, like that in the realistic heavy-ion collision experiments, it is necessary to discuss $\Delta$'s behaviors on the freezeout line. A hypothetical freezeout line is chosen for the following calculations because the relative position of the freezeout line to the phase transition line is not known to us.  The evolution trajectories are chosen to be a series of vertical lines with constant chemical potential, and the magnitude of chemical potential $\mu$ ranges from $200$ MeV to $280$ MeV. For different trajectories, the initial temperature is set to be $10.1$ MeV higher than the phase transition temperature, and the freezeout temperature is set to be $13.5$ MeV lower than the phase transition temperature.
The decreasing of temperature and the expansion of volume is set again as that in Eq.\,(\ref{temp}) and (\ref{volume}), and the initial volume $V_0=8^3 \,\rm{fm}^3$.

\begin{figure}[tpb]
\centering
\includegraphics[width=0.8\columnwidth]{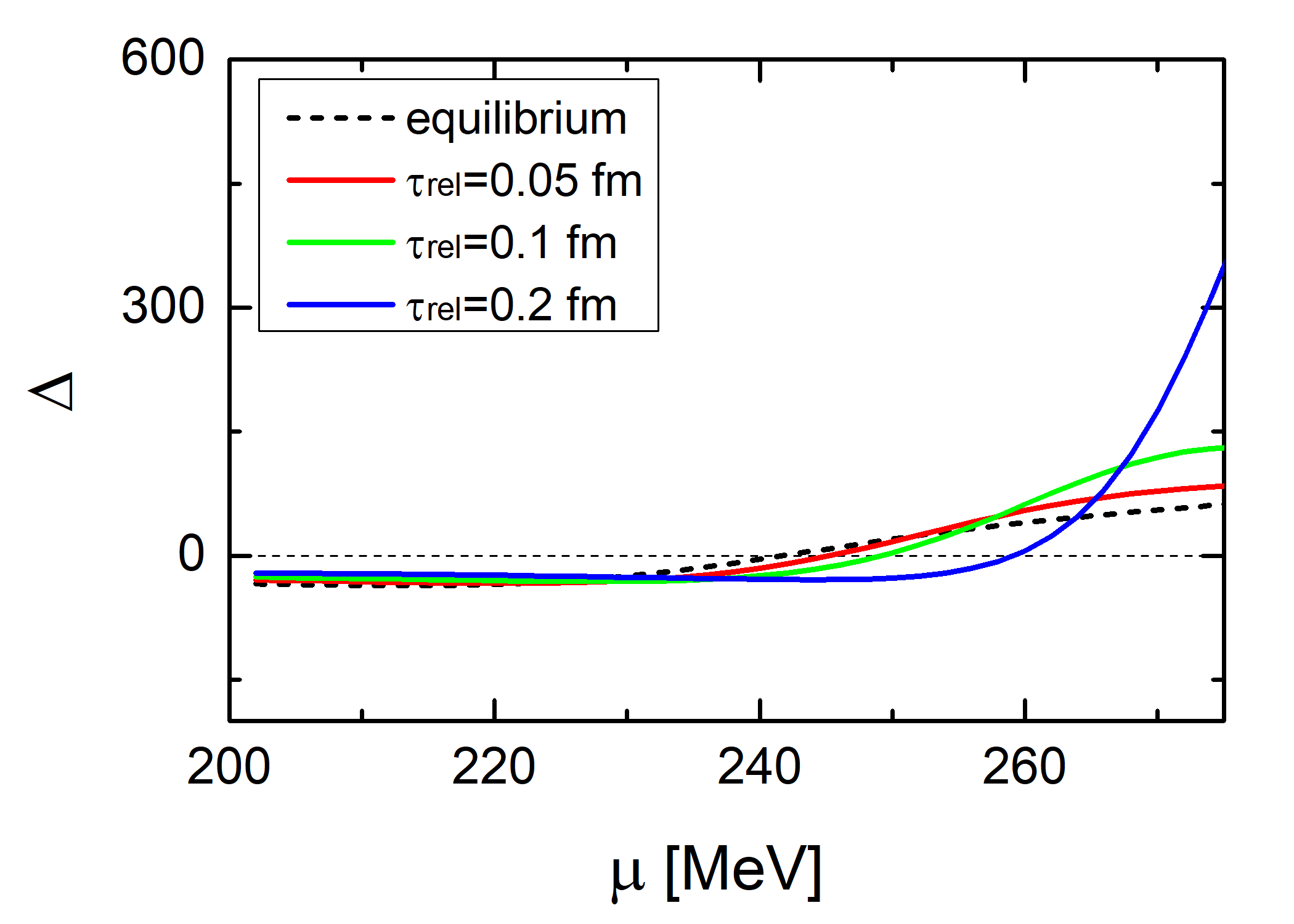}
\vspace*{-3mm}
\caption{(color online) The dynamical results of $\Delta$ (the colored lines) and the equilibrium result (the dashed lines) along a hypothetical freeze-out line. The different colored lines are for the results with initial relaxation rate $\tau_{rel} = 0.05, 0.1, 0.2\,$fm, respectively.}
\label{delta-freeze}
\end{figure}

In Fig.\,\ref{delta-freeze}, we present the numerical results of $\Delta$ on the freezeout line, the colored lines are results of dynamical $\Delta$ with the initial relaxation rate $\tau_{rel}=0.05, 0.1, 0.2$ fm respectively, the dashed line is the equilibrium result for comparison.
With the current setup and the choice of the freezeout line, we find that the sign of $\Delta$  turns to positive at large chemical potentials for both the equilibrium and dynamical $\Delta$. With a larger relaxation rate, $\Delta$ turns to positive at a larger $\mu$.
From the subfigure with $\mu=260$ MeV and  $\mu=270$ MeV in Fig.\,\ref{delta-mu}, we can see that there are large positive $\Delta$ regions below the phase transition temperature. If the system freezes out in this region, the positive signal of $\Delta$ is possibly observed after the dynamical evolution. But note that there is also the possibility that the realistic freezeout line lies in the negative $\Delta$ region, in this case, the first-order phase transition signal could not be observed. Emphasize again that the experimental observables are also affected by other factors, like the magnitude of the relaxation rate, the size of the system, the evolution trajectories, and other noncritical factors. In the current stage, our calculations can only provide reference information for the experimental detection.

\section{Summary and outlook}
\label{sec4}

To summarize, we study the dynamical behaviors of the first-order phase transition criterion in a finite-size system. The thermodynamic system is described by a parameterized Landau free energy in the $\mu$-$T$ plane.
A dynamical framework for the Landau free energy and the first-order phase transition criterion is established based on the Fokker-Planck equation.
By setting the evolution trajectories to be along the constant chemical potential lines and the system volume to be finite,
we analyze separately the dynamical behaviors of $\Delta$ in different phase transition scenarios and find that the positive $\Delta$ signal is preserved under certain conditions, which is also delayed in the dynamical process.
The positive $\Delta$ signals that only appear at the first-order phase transition region with $\mu >\mu_{c}$, are strongly influenced by the finite-size effects and the dynamical effects.
In detail, the temperature intervals for the positive $\Delta$ signals are shifted to the lower temperature region compared with the equilibrium case.
However, analyzing the volume size, the relaxation rate, and the initial temperature's influence on the $\Delta$ shows that a large volume size,
a large relaxation rate or a higher initial temperature away from the phase transition temperature will all suppress the dynamical signal of $\Delta$.
In addition, the $\Delta$ signals along curved trajectories are discussed and we find that the $\Delta$ signal is pulled down when the trajectories are bent over to the crossover side.
Further analyses with expanding volume show that $\Delta$ presents qualitatively similar behaviors as the former discussions with comparable and fixed volume, and the positive $\Delta$ signal is possibly preserved on the hypothetical freezeout line in the large chemical potential region.

Note that as a first step to studying the dynamical behaviors of the first-order phase transition signal, the setup of the system is simplified for the calculations and discussions, which, on the other hand, push it away from the realistic phase transition that occurs in heavy ion collisions.
Improvements such as extending the parametrization of free energy to the sixth power make it possible for us to discuss the dynamics of hyper-order cumulants and improve the calculations of $\Delta$, which will be done in the near future.
In addition, more efforts like embedding the $\Delta$ signal in the hydrodynamic equations are urgently needed so that a more realistic experimental process can be simulated.
The mixture of the first-order phase transition signal with the other factors unrelated to the phase transition also needs further study.\\
\\
\appendix{\textbf{Appendix: More results with expanding volume}}

\begin{figure*}[tbp]
\centering
\includegraphics[width=2.0\columnwidth]{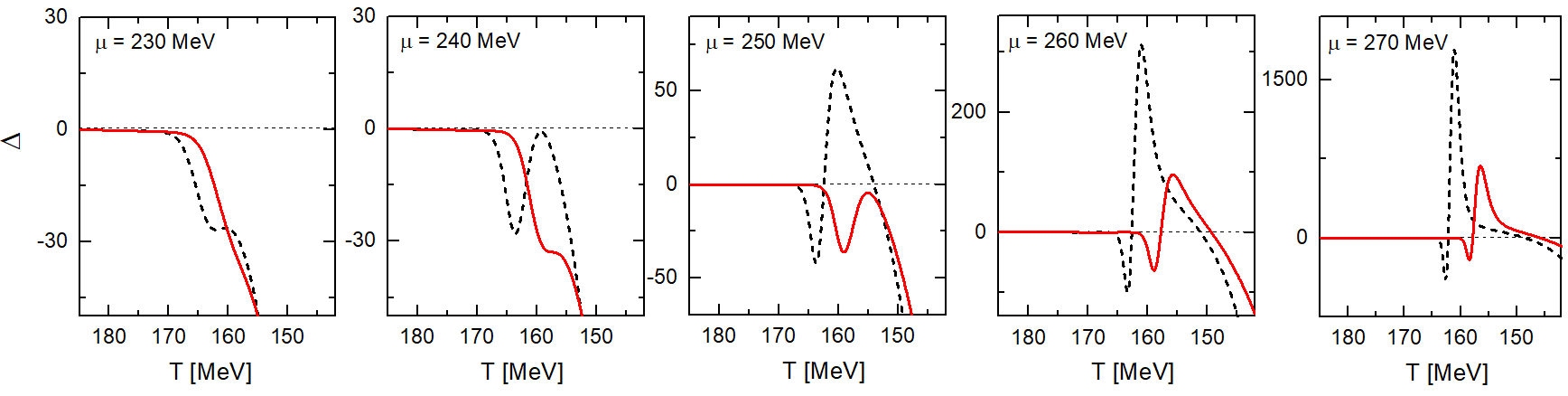}
%\vspace*{-6mm}
\caption{(color online) The equilibrium  and  dynamical $\Delta$ as functions of the temperature $T$, with the initial system volume $V_0 = 8^{3}\,\textrm{fm}^{3}$ and the initial relaxation rate $\tau_{rel} = 0.1\,$fm. The dashed lines represent the instantaneous equilibrium results and the red solid lines represent the dynamical results. In different subfigures, the evolution trajectories are set to be along different constant $\mu$ lines in the $\mu$-$T$ plane. }
\label{delta-mu-Vt}
\end{figure*}

In this appendix, by setting the system volume to expand in a Hubble-like way (defined in Eq.\,(\ref{volume})), the initial volume $V_0=8^3$ fm$^3$, we first present the numerical results of $\Delta$ along vertical trajectories with different constant chemical potential in Fig.\,\ref{delta-mu-Vt}, thus the system undergoes different phase transition scenarios. Similar to the results in Fig.\,\ref{delta-mu}, the positive $\Delta$ signal appears only when the chemical potential is large enough,  corresponds to a wider coexistence region for different phases during the evolution.

\begin{figure*}[tbp]
\centering
\includegraphics[width=1.5\columnwidth]{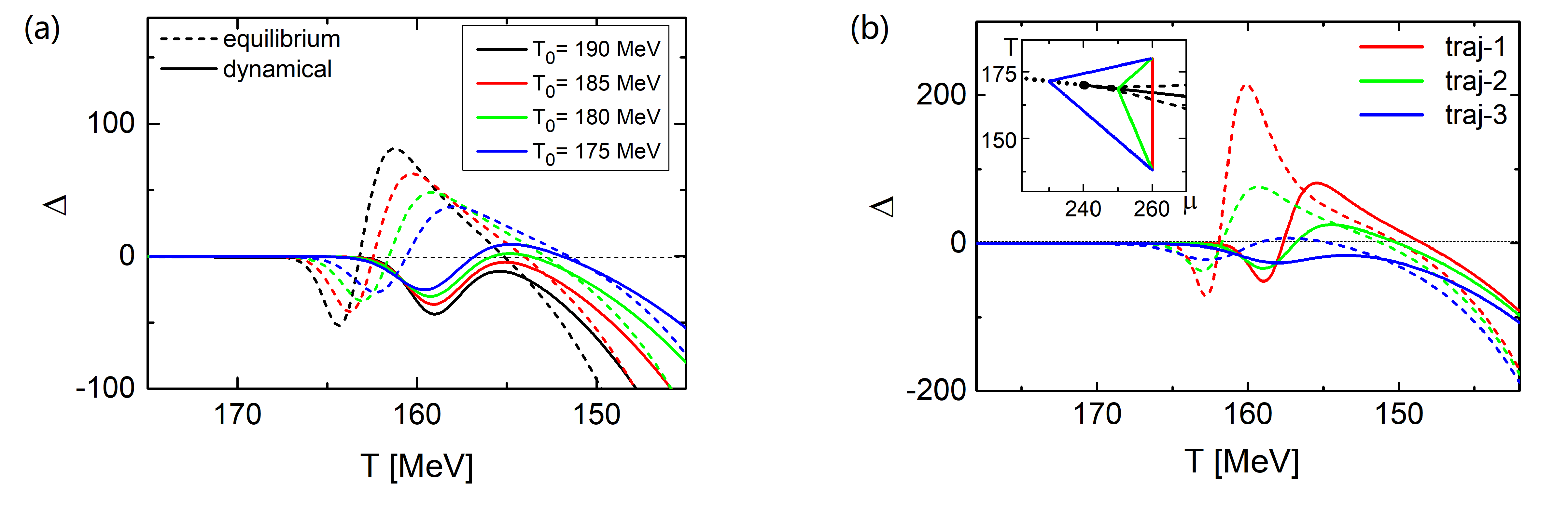}
\vspace*{-3mm}
\caption{(color online) (a) The dynamical $\Delta$ (the colored lines) as functions of the temperature with initial temperature $T_{0}=190, 185, 180, 175\;$MeV respectively. The baryon chemical potential is set at $\mu=250\;$MeV. (b) The dynamical  $\Delta$ along the three evolution trajectories in the $\mu$-$T$ plane shown in the inset. In both subfigures, the dashed lines are the equilibrium results for comparison, the initial system volume $V_0 = 8^{3}\,\textrm{fm}^{3}$, the initial relaxation rate is $\tau_{rel}=0.1\,$fm, and
the initial system volume $V_0 = 8^{3}\,\textrm{fm}^{3}$. }
\label{delta-T0-Vt}
\end{figure*}

To make comparisons with the results in Fig.\,\ref{delta-T0}, in the subfigure (a) of Fig.\,\ref{delta-T0-Vt}, we present calculations on $\Delta$ along the vertical trajectory $\mu=250$ MeV, but with different initial temperatures. The colored lines are for $\Delta$ with initial temperature $T_{0}=190, 185, 180, 175$ MeV respectively, and  the dashed lines are for the equilibrium $\Delta$. Note that at the initial time, the system volume is the same for different $T_{0}$, it is then different when the temperature is around $160$ MeV, which leads to the difference of the equilibrium $\Delta$ in different $T_{0}$ cases. But qualitatively, the temperature interval for positive $\Delta$ signal is more evident as the initial temperature is closer to the phase transition region, which is similar to that in Fig.\,\ref{delta-T0}.
In the subfigure (b) of Fig.\,\ref{delta-T0-Vt}, we also make calculations on $\Delta$ along curved trajectories, to compare with the results in Fig.\,\ref{delta-3trajs}. Except the volume size, the other parameters are the same as that in Fig.\,\ref{delta-3trajs}. The positive $\Delta$ signal is relatively obvious along traj-1, but is suppressed for the other two trajectories, which the coexistence region takes less proportion during the evolution.

\section*{Acknowledgments}
We thank Huichao Song, Yi Lu, Baochi Fu and Jun-Hui Zheng for fruitful discussions.
LJ acknowledges the support from the National Natural Science Foundation of China (NSFC) under grant No. 12105223 and No.12247103,
FG is supported by the National  Science Foundation of China under Grants  No. 12305134,
and YL acknowledges the support from the NSFC under Grant No. 12175007 and No. 12247107.

\end{document}